\documentclass[pra,twocolumn,showpacs]{revtex4}
\usepackage{epsfig}
\usepackage{amsbsy}
\usepackage{amsmath}
\usepackage{latexsym}

\begin{document}

\title{Boundary and impurity effects on entanglement of Heisenberg chains }
\author{Xiaoguang Wang}
\affiliation{Australian Centre of Excellence for Quantum Computer Technology, Department of Physics,	\\
Macquarie University, Sydney, New South Wales 2109, Australia.}

\date{\today}
\begin{abstract}
We study entanglement of a pair of qubits and the bipartite entanglement between the pair and the rest within open-ended Heisenberg $XXX$ and $XY$ models. The open boundary condition leads to strong oscillations of entanglements with a two-site period, and the two kinds of entanglements are 180 degree out of phase with each other. The mean pairwise entanglement and ground-state energy per site in the $XXX$ model are found to be proportional to each other. We study the effects of a single bulk impurity on entanglement, and find that there exists threshold values of the relative coupling strength between the impurity and its nearest neighbours, after which the impurity becomes pairwise entangled with its nearest neighbours.
\end{abstract}
\maketitle

\section{Introduction}
Over the past few years much effort has been put into studying quantum entanglement in various quantum spin models. At zero temperature, entanglement naturally exists in many-body ground states. The systems contain two classes, finite spin clusters~\cite{Nielsenphd,OConnor,Vedral1,Wang,Vedral2,WangFu,Kamta,Yeo,Victory,ChenHong,Zhou,Saguia,Schliemann,Meyer,Syljuasen,Santos,Glaser,TMeyer} such as a ring of qubits, and infinite spin systems~\cite{Osborne,Fazio,Vidal,Bose,Vidall,Gu,Osenda,Korepin} where quantum phase transition~\cite{QPT} may occur. For a ring of qubits interacting via the Heisenberg $XXX$ Hamiltonian, the concurrence for the ground state, quantifying  entanglement of a pair of qubits, is equal to the absolute value of the ground-state energy per site~\cite{OConnor,Victory}. Universal scaling behaviours of quantum entanglement of ground states emerge at a quantum phase transition point in the anisotropic $XY$ model~\cite{Osborne,Fazio,Vidal}.  
Experimentally, entangled state of magnetic dipoles has been found be to crucial to describing magnetic behaviours in a quantum spin system~\cite{Coppersmith}.

In most of the previous studies on entanglement of many-body ground states, a periodic boundary condition (PBC) is assumed for spin chains. In compounds such as Mn$_{12}$ and Fe$_8$, the metal ions within a single molecule form almost a perfect ring, which can be described by the Heisenberg interaction~\cite{Waldmann}.  
On the other hand, spin chains with an open boundary condition (OBC) have been used to construct spin cluster qubits~\cite{Loss,Loss2} for quantum computation and  employed for quantum communication from one end to another~\cite{Sougato,Ekert}. Perfect state transfer have been obtained via the open chain without requiring qubit coupling to be switched on and off~\cite{Ekert}. These investigations reveal that open chains are of great advantage in implementing quantum information tasks. Thus, the study of the entanglement structure in open spin chains will be of importance as the entanglement underlies operations of quantum computing and quantum information processing.  Here, we investigate the open boundary effects on ground-state entanglement.

A linear open $N$-qubit chain can be viewed as a ring of $(N+1)$-qubit chains with an impurity (($N+1$)-th qubit), where the coupling $J_{N+1,1}=J_{N+1,N}=0$.  
The impurity plays a very important role in condensed matter physics, and it will be interesting see its effect on quantum entanglement.
In earlier studies, impurity effects on entanglement have been considered with a small three-spin system~\cite{Impurity}. Here, we consider large spin systems described by the Heisenberg $XY$ model. 

The paper is organized as follows. In Sec.~IIA, by exact diagonlization method, we study the ground-state entanglement in the Heisenberg $XXX$ model with an OBC.
The mean nearest-neighbour entanglement is found be proportional to the ground-state energy per site. For studying systems with large number of site, in Sec.~IIB, we consider the $XY$ model with an OBC, which can be solved exactly by Jordan-Wigner mapping~\cite{JW}. Entanglements display strong oscillations over lattice. With increase of the ground-state energy, the mean pairwise entanglement increases in the $XY$ model, while decreases in the $XXX$ model.
In Sec.~III, we study impurity effects on entanglement of the spin systems described by the $XY$ model, where the impurity is located in the bulk of the system. We conclude in Sec.~IV.

\section{Boundary effects}
\subsection{Heisenberg $XXX$ model}

We consider a physical model of $N$ qubits interacting via the isotropic Heisenberg Hamiltonian with an OBC,
\begin{equation}
H_{XXX}=\frac{J}{2}\sum_{i=1}^{L-1}\big(1+
\sigma_{ix}\sigma_{i+1x}+\sigma_ {iy}\sigma_{i+1y}+\sigma_{iz}\sigma_{i+1z}
\big),  \label{h}
\end{equation}
where $\vec{\sigma}_i=(\sigma _{ix},\sigma _{iy},\sigma _{iz})$ is the
vector of Pauli matrices and $J$ is the exchange constants. The positive and
negative $J$ correspond to the antiferromagnetic (AFM) and ferromagnetic
(FM) case, respectively. 

The SU(2) symmetry ($[H,S_\alpha]=0)$ is evident, where $S_\alpha=\sum_{i=1}^N \sigma_{i\alpha}/2, \alpha\in\{x,y,z\}$. This symmetry guarantees that reduced density matrix $\rho_{ij}$ of two qubits, say qubit $i$ and $j$, for the ground state $\rho$ has the form~\cite{OConnor} 
\begin{equation}
\rho _{ij}=\left( 
\begin{array}{llll}
u_{ij} & 0 & 0 & 0 \\ 
0& w_{ij} & z_{ij} &0  \\ 
0& z_{ij} & w_{ij} & 0 \\ 
0& 0 & 0 & u_{ij}
\end{array}
\right)   \label{eq:rho12}
\end{equation}
in the standard basis $\{|00\rangle ,|01\rangle ,|10\rangle ,|11\rangle \}.$

From the reduced density matrix, the concurrence~\cite{Conc} quantifying the
entanglement is readily obtained as~\cite{OConnor,Victory} 
\begin{align}
C_{i,j}=&2\max \left[0,|z_{ij}|-u_{ij}\right]  \nonumber\\
       =&\max \left[0,|G_{i,j}^{xx}|-G_{i,j}^{zz}/2-1/2\right]\nonumber\\
=&\max\left(0,|G_{ij}^{zz}|-G_{ij}^{zz}/2-1/2\right)\nonumber\\
=&\max\left(0, -3G_{ij}^{zz}/2-1/2\right),
\label{ccc}
\end{align}
where $G_{ij}^{\alpha\alpha}=\text{Tr}(\sigma_{1\alpha}\sigma_{2\alpha}\rho)$
are correlation functions.
The second equality follows from the relations between matrix elements of $\rho_{ij}$ and correlation functions given by $u_{ij}=(1+G^{zz}_{ij})/4 $ and $z_{ij}=G_{ij}^{xx}/2$. The SU(2) symmetry lead to the third equality. The fourth equality follows from the inequality $|G^{zz}_{ij}|\le 1$, which is a 
special case of a more general result that $|\langle A\rangle|\le 1$ for any Hermitian operator $A$ satisfying $A^2=1$. We see that the qubit $i$ and $j$ are entangled if $G_{ij}^{zz}<-1/3$.

For studying bipartite entanglement, we consider even number of sites, for which ground states of the Heisenberg model are nondegenerate, and thus pure. The bipartite entanglement of pure states is well-defined by entropies of one subsystem. 
Here, we choose the linear entropy to quantify the bipartite entanglement between a pair of qubits and the rest. {}From Eq.~(\ref{eq:rho12}), in a similar way to get Eq.~(\ref{ccc}), the linear entropy for $\rho_{ij}$ is obtained as
\begin{equation}
E_{ij}=1-\text{Tr}(\rho_{ij}^2)=1-\frac{1}{4}\left[1+3(G^{zz}_{ij})^2\right]. \label{eee}
\end{equation}
The reduced density matrix $\rho_{ij}$ is only determined by the correlation function $G^{zz}_{ij}$, so do the concurrence and the linear entropy.

\begin{figure}
\includegraphics[width=0.45\textwidth]{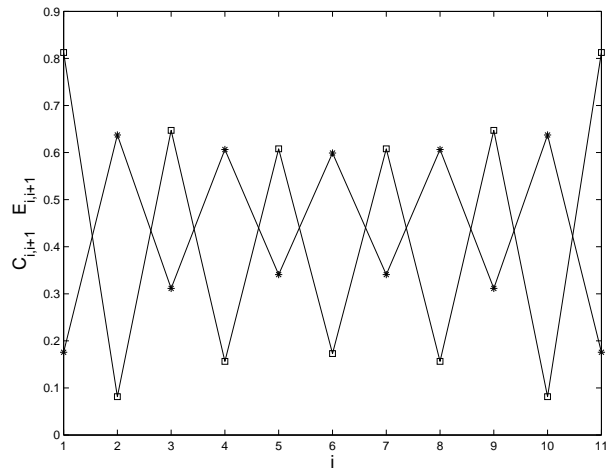}
\caption{Ground-state nearest-neighbour concurrence (square line) and linear entropy (star line) versus site number for $L=12$ in the Heisenberg $XXX$ model. }
\end{figure}

Due to the nearest-neighbour nature of the interaction in our system, the entanglement between a pair of nearest qubits is expected to be prominent comparing with a pair of non-nearest-neighbour qubits. Thus, we focus on the nearest-neighbour case in the following discussions. For the isotropic Heisenberg Hamitonian with a PBC, entanglements between 
qubit $i$  and $i+1$ are independent on index $i$. However, for the case of OBC, the concurrence $C_{ii+1}$ and the linear entropy $E_{ii+1}$ must be site-dependent due to the breaking of translational symmetry.  Next, we study entanglements using exact diagonalization method.

We first diagonalize Hamitonian $H_{XXX}$ to obtain the ground state, from which we calculate the correlation function $G_{ii+1}^{zz}$, and then the pairwise and bipartite entanglements via Eqs.~(\ref{ccc}) and (\ref{eee}), respectively. 
The numerical results for 12 qubits (4069-dimensional Hibert space) are shown in Fig.~1. The entanglements oscillate with a two-site period, and the concurrence and the linear entropy are 180 degree out of phase with each other. The pair with qubits 1 and 2, and the one with qubits $L-1$ and $L$ display maximal pairwise entanglement (minimal bipartite entanglement). For two qubits interacting via the Heisenberg Hamiltonian, they prefer to the maximally entangled singlet state $|\Psi\rangle=1/\sqrt{2}(|01\rangle-|10\rangle)$ with concurrence $C=1$. Qubit 1 favor maximal entanglement with the only nearest-neighbour qubit 2, and result in higher concurrence $C_{12}$. For qubit 2, there are a competition between qubit 1 and 3, and they both favor maximally entangled with qubit 2. Qubit 2 shares large entanglement with qubit 1, and thus results in less entanglement with qubit 3, and the oscillatory feature appears. As the pair of qubits 1 and 2 have strong entanglement, the bipartite entanglement between the pair of qubits with the rest of the system is relatively suppressed as we have seen from the figure.  

It will be instructive to give two extreme examples to reveal relations between pairwise entanglement and the biparite entangelement between the pair and the rest. The first one is the singlet state with $C=1$ and $E=0$, and second one is the maximally mixed state $\rho_{12}=\text{diag}(1/4,1/4,1/4,1/4)$ with $C=0$ and $E=0.75$ (maximal linear entropy). Thus, in general, the above numerical results and the two simple examples suggest that the more the pairwise entanglement, the less the bipartite entanglement. Of course, this is not a general statement for arbitrary multi-qubit states.

\begin{figure}
\includegraphics[width=0.45\textwidth]{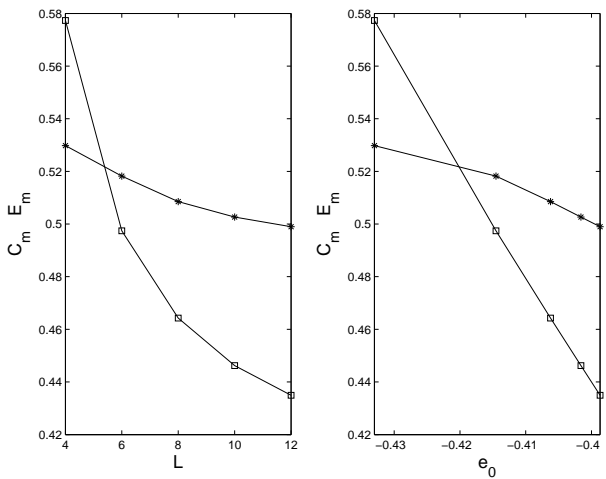}
\caption{Mean concurrence (square line) and mean linear entropy (star line) versus $L$ and $e_0$ in the Heisenberg $XXX$ model. }
\end{figure}

Next, we consider the mean entanglement and study its relations with ground-state energy ${\cal E}_0$. The nearest-neighbour mean concurrence and linear entropy can be defined as
\begin{align}
C_m(L)&\equiv\frac{1}{L-1}\sum_{i=1}^{L-1} C_{i,i+1}, \nonumber\\
E_m(L)&\equiv\frac{1}{L-1}\sum_{i=1}^{L-1} E_{i,i+1}.
\end{align}  
We made numerical calculations for even $L$ from 4 to 12, and the entanglements versus $L$ or $e_0$ are shown in Fig.~2, where $e_0={\cal E}_0/L$ is the ground-state energy per site. The pairwise entanglement decreases as $L$ or $e_0$ increases, and so does the bipartite entanglement. More interestingly, the pairwise entanglement decreases linearly with $e_0$. We may obtain an analytical relation between $C_m$ and $e_0$ as follows. 

The numerical results for the finite lattices show that $C_{ii+1}>0$, and thus from Eq.~(\ref{ccc}) we obtain
\begin{equation}
G_{ii+1}^{zz}= \frac{-2C_{ii+1}-1}{3}. 
\label{cccccc}
\end{equation}
From $H_{XXX}$~(\ref{h}) and after taking into account the SU(2) symmetry, we get
\begin{equation}
e_0=\frac{L-1}{2L}+\frac{3}{2L}\sum_{i=1}^{L-1}G_{ii+1}^{zz}. \label{eeee}
\end{equation}
Substituting Eq.~(\ref{cccccc}) into (\ref{eeee}) leads to 
\begin{equation}
e_0=-\frac{L-1}{L}C_m(L),
\end{equation}
where we have used the definition of mean concurrence. 
The above relation shows that the mean entanglement is proportional to the ground-state energy per site, implying that less energy gives more pairwise entanglement. Although this relation is obtained for finite lattices, we make a conjecture that it is valid for any number of lattice sites. 

The concurrence and linear entropy obtained for finite sites of lattice up to 12 are not so close to those for infinite lattice given by
\begin{align}
C_{m}(\infty)=&2\ln 2-1\approx 0.3863, \nonumber\\ 
E_{m}(\infty)=&\frac{2}{3}[1-\ln 2 (2\ln 2-1)]\approx 0.4882.
\end{align}
This is due to the finiteness of lattice, and also due to the effects of boundaries. To investigate effects of open boundaries on entanglement in {\em larger} systems, we next study quantum Heisenberg $XY$ model, which can be solved exactly by the Jordan-Wigner transformation~\cite{JW}.

\subsection{Heisenberg $XY$ model}
The Heisenberg $XY$ Hamiltonian with an OBC is given by
\begin{equation}
H_{XY}=\frac{J}2\sum_{i=1}^{N-1}\big( \sigma_{ix}\sigma_{i+1x}+\sigma_ {iy}\sigma_{i+1y}\big).
\end{equation}
There are two symmetries in the Hamiltonian. One is a U(1) symmetry ($[H,S_z]=0)$, and another is a $Z_2$ symmetry ($[H,\Sigma_x]=0$), 
where $\Sigma_x=\sigma_{1x}\otimes\sigma_{2x}\otimes\cdots\otimes\sigma_{Nx}$.
They guarantee that reduced density
matrix $\rho_{ij}$ of two qubits $i$ and $j$ for the ground state $\rho$ is given by Eq.~(\ref{eq:rho12}).
For the lack of $SU(2)$ symmetry, the correlation functions $G_{ij}^{zz}$ and 
$G_{ij}^{xx}$ are no longer equal, and then the concurrence is given by
\begin{equation}
C_{ij}=\max \left[0,|G_{ij}^{xx}|-G_{ij}^{zz}/2-1/2\right].
\label{cccc}
\end{equation}
determined by two correlation functions.

For nearest-neighbor cases, two correlation functions $G_{ii+1}^{xx}$ and $G_{ii+1}^{zz}$ are dependent, and the latter can be written in terms of the former as we will see shortly, and thus the concurrence is only determined by a single correction function $G_{ii+1}^{xx}$. To show this, we use the well-know Jordan-Wigner mapping~\cite{JW}, under which Hamiltonian $H_{XY}$ {\em exactly} maps to 
\begin{equation}
H_{XY}=J\sum_{i=1}^{L-1}\big(c_i^{\dagger }c_{i+1}+c_{i+1}^{\dagger }c_i\big)
\end{equation}
due to the OBC, where $c_i$ and $c_i^\dagger$ are fermionic annihilation and creation operators, respectively. 
Then, in the fermionic representation, the two correlation functions can be written as
\begin{align}
z_{ii+1}=&\langle c_i^\dagger c_{i+1}\rangle,\nonumber\\
u_{ii+1}=&
\langle c_i^\dagger c_{i} c_{i+1}^\dagger c_{i+1}\rangle
=1/4- |\langle c_i^\dagger c_{i+1}\rangle |^2, \label{ele}
\end{align}
where we have used Wick theorem~\cite{Wick}, and the fact $\langle c_i^\dagger c_i\rangle=1/2$ for any $i$. Thus, the concurrence is determined only by one correlation function 
$\langle c_i^\dagger c_{i+1}\rangle$,
\begin{equation}
C_{ii+1}=2\max(0, |\langle c_i^\dagger c_{i+1}\rangle|+\langle c_i^\dagger c_{i+1}\rangle^2-1/4)\label{newc}
\end{equation}

To compute the correlation function $\langle c_i^\dagger c_{i+1}\rangle$, we first diagonalize Hamitonian $H_{XY}$ by the following transformation
\begin{equation}
c_n=\sum_{k=1}^Lg_{nk}\tilde{c}_k=\sqrt{\frac 2{L+1}}\sum_{k=1}^L\sin \left( 
\frac{nk\pi }{L+1}\right) \tilde{c}_k
\end{equation}
After the transformation, Hamiltonian $H_{XY}$ can be written as
\begin{equation}
H_{XY}=\sum_{k=1}^L\epsilon _k\tilde{c}_k^{\dagger }\tilde{c}_k=\sum_{k=1}^L -2%
\cos \left( \frac{k\pi }{L+1}\right) \tilde{c}_k^{\dagger }\tilde{c}_k,
\end{equation}
where $\epsilon_i$ is the single-particle energy, and $J=-1$ is assumed for convenience. The ground state is then expressed as 
\begin{equation}
|\Psi\rangle_{GS} =\tilde{c}_{_1}^{\dagger}\tilde{c}_{2}^{\dagger
},...,\tilde{c}_{L/2}^{\dagger}|0\rangle,
\end{equation}
with energy ${\cal E}_{0}=\sum_{k=1}^{L/2}\epsilon
_{k}$. From the above expression of the ground state, the correlation function is obtained as 
\begin{equation}
\langle c_i^{\dagger }c_{i+1}\rangle =\sum_{k=1}^{L/2}g_{i,k}g_{i+1,k}.
\label{corre}
\end{equation}
The combination of Eqs.~(\ref{newc}) and (\ref{corre}) gives exact analytical expression for the concurrence. By using 
Eq.~(\ref{ele}), the linear entropy of the two-qubit reduced density matrix $\rho_{ii+1}$ is obtained as
\begin{equation}
E_{ii+1}=3/4-4|\langle c_i^{\dagger }c_{i+1}\rangle |^4-2|\langle c_i^{\dagger }c_{i+1}\rangle |^2,
\end{equation}
which is also analytical. From these analytical expressions, entanglements are readily calculated numerically for a very large $L$.

\begin{figure}
\includegraphics[width=0.45\textwidth]{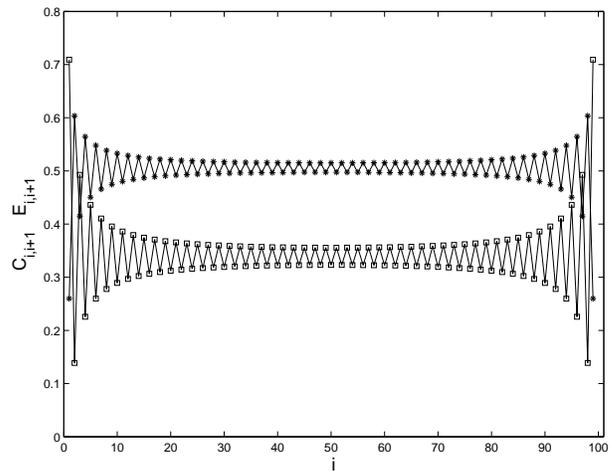}
\caption{Ground-state nearest-neighbour concurrence (square line) and linear entropy (circle line) entanglement versus site number for $L=100$ in the Heisenberg $XY$ model. }
\end{figure}

Figure 3 displays features of entanglement distribution over lattice sites, which are similar to those of the Heisenberg $XXX$ model. Strong oscillations of entanglements are more evident. The similarity arises since two qubits interacting via the Heisenberg $XY$ interaction also favor the singlet state. Near the bulk area, the entanglements oscillate with nearly the same amplitude with respect to the mean value, and the boundary effects diminish. 

Now, we study the mean entanglements for different $L$. 
Figure 4 gives the entanglement versus $L$ and the energy per site.
With the increase of $L$, the mean concurrence decreases, while the linear entropy increases, and both saturate for large $L$. For a infinite lattice, the concurrence and the linear entropy are given by
\begin{align}
C_m(\infty)=&\frac{2}{\pi}+\frac{2}{\pi^2}-\frac{1}2\approx 0.3393,\nonumber\\
E_m(\infty)=&\frac{3}4-\frac{4}{\pi^4}-\frac{2}{\pi^2}\approx 0.5063.
\end{align}
We see that the concurrence and linear entropy with $L=200$ are nearly identical to those with infinite $L$. 
From Fig.~4, we observe that the mean concurrence increases, while the mean linear entropy decreases with the increase of energy. The behaviour of the mean concurrence is opposite to that in the Heisenberg $XXX$ model. Here, the lower energy does not correspond to higher pairwise entanglement.

\begin{figure}
\includegraphics[width=0.45\textwidth]{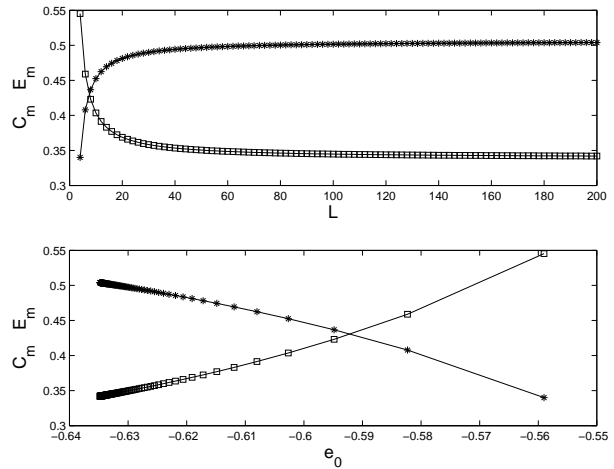}
\caption{Mean concurrence (square line) and mean linear entropy (star line) versus $L$ and $e_0$ in the Heisenberg $XY$ model. }
\end{figure}

\section{Impurity effects}
As we have stated in the introduction, a linear open $N$-qubit chain can be viewed as a ring of $(N+1)$-qubit chains with an impurity.
In this section, we study single impurity effects on entanglement of Heisenberg chains. We consider the Heisenberg $XY$ model described by the Hamiltonian
\begin{equation}
H_{XY}=\sum_{i=1}^{N-1}\frac{J_{i,i+1}}{2}\big(\sigma_{ix}\sigma_{i+1x}+\sigma_ {iy}\sigma_{i+1y}\big).
\end{equation}
with an OBC. Now, we assume the single impurity spin is located in the bulk (site $N/2$)~\cite{Stolze}, namely,
\begin{align}
J_{N/2-1,N/2}&=J_{N/2,N/2+1}=J'=\alpha J, \nonumber\\
J_{i,i+1}&=J=1 \,\text{for other} \; i,
\end{align}
where $\alpha$ characterizes the relative strength of the coupling between the impurity qubit and its nearest neighbours. To be specific, we focus on pairwise entanglement in the following discussions.

Due to the existence of a single impurity in the bulk, we do not expect  analytical results for entanglement. However, the Hamiltonian still have the 
U(1) and $Z_2$ symmetries, we can map the Hamiltonian exactly to a fermionic one, and the Wick theorem applys. Thus, Eqs.~(\ref{cccc}) and (\ref{newc}) for the concurrence are  valid and applicable to the impurity model. All we need to is to diagonalize a $L\times L$ matrix to obtain the coefficients  $g_{nk}$ in the expression $c_n=\sum_{k=1}^L g_{nk}\tilde{c}_k$, and then calculate the correlation function $\langle c_{i}^\dagger c_{i+1}\rangle $ via the relation $\langle c_{i}^\dagger c_{i+1}\rangle=\sum_{k=1}^{L/2}g_{ik}^* g_{i+1,k}$. Note that the summation is from $k=1$ to $L/2$ since the system considered is non-degenerate (the single-particle energy $\epsilon_k$ is not zero) and number of the negative values of $\epsilon_k$ are $L/2$. 

From Fig.~5, we see that the impurity leads to additional oscillations of the pariwise entanglement in the bulk region.
For small $\alpha=0.1$, the concurrences $C_{N/2-1,N/2}$ and 
$C_{N/2,N/2+1}$ are zero. Even the coupling between impurity qubit and its nearest neighbours are not zero, the entanglement between them vanishes due to competition among qubits. For larger $\alpha=0.6$, the entanglements between impurity qubit and qubits $L/2+1$ and $L/2-1$ build up, where $C_{N/2-1,N/2}>C_{N/2,N/2-1}$, which also holds for  $\alpha=1.2>1$ and $\alpha=2>1$. The difference between the two concurrences results from the choice of even $N$, leading to a non-symmetry of the entanglement distribution. We see that even one single impurity have strong effects on entanglement structure, especially in the region near the impurity.

From the above analyses, it is expected there exists a threshold value of $\alpha$, after which the impurity qubit and its nearest neighbors becomes entangled. In Fig.~6, we plot the nearest-neighbour concurrences in the bulk region as a function of $\alpha$. It is evident that there exist threshold values $\alpha_{1}$ for concurrence $C_{N/2-1,N/2}$ and $\alpha_{2}$ for concurrence $C_{N/2,N/2+1}$. The threshold value 
$\alpha_{2}$ is slightly larger than $\alpha_{1}$, and the pairwise entanglement between the impurity and qubit $N/2-1$ is always a little stronger than that between the impurity and qubit $N/2+1$ when $\alpha>\alpha_2$. We also observe that the increase of $\alpha$ suppresses the concurrence $C_{N/2+1,N/2+2}$, while enhance the concurrence $C_{N/2+2,N/2+3}$.
When $\alpha=1$ (no impurity case), all concurrences shown in the figure are almost identical since we have chose a large $L=500$, which diminishes the boundary effects in the bulk and the amplitudes of oscillations become very small.

\begin{figure}
\includegraphics[width=0.45\textwidth]{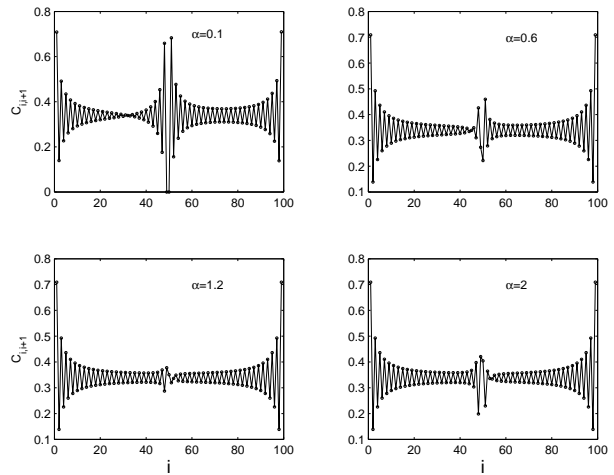}
\caption{Ground-state nearest-neighbour concurrence versus site number for $L=100$ in the Heisenberg $XY$ model with an impurity. }
\end{figure}

\begin{figure}
\includegraphics[width=0.45\textwidth]{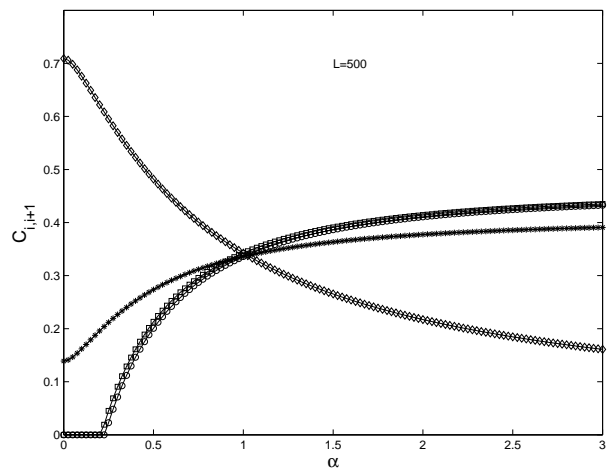}
\caption{ Ground-state nearest-neighbour concurrences $C_{N/2-1,N/2}$ (square line), $C_{N/2,N/2+1}$ (circle line), $C_{N/2+1,N/2+2}$ (diamond line), and $C_{N/2+2,N/2+3}$ (star line) as a function of $\alpha$ for the impurity model. }
\end{figure}

\section{Conclusion}
In conclusion, we have studied ground-state entanglements in Heisenberg $XXX$ and $XY$ models with an OBC. The OBC leads to strong oscillations with a two-site period of entanglement in a pair of nearest-neighbour qubits and bipartite entanglement between the pair and the rest. The maximal pairwise entanglement and minimal bipartite entanglement occurs at open ends, and the two kinds of entanglements are 180 degree out of phase with each other. 
In both models, the two-qubit reduced density matrix is determined by only one correlation function, and so do the entanglements. We have found that the mean entanglement is proportional to the ground-state energy per site in the $XXX$ model. With increase of the ground-state energy, the mean pairwise entanglement decreases in the $XXX$ model, while increases in the $XY$ model. 

We study the effects of a single bulk impurity on entanglement, and find that 
the impurity leads to additional oscillations of entanglement in the bulk region. We also find that there exists threshold values of the relative coupling strength between the impurity and its nearest neighbours, after which the impurity becomes entangled with its nearest neighbours. As the entanglement underlies operations of quantum computation and quantum information processing, 
the structures of entanglement found in the present studies are useful when we make a simulation of quantum systems where boundary and impurity effects cannot be negligible. 

\acknowledgments This work has been supported by an Australian 
Research Council Large Grant and Macquarie University Research Fellowship. We thanks for the helpful discussions with Dr. S.J. Gu and Prof. Barry C Sanders.

\end{document}